\documentstyle[12pt]{article}

\def\reference{\parskip 0pt\par\noindent\hangindent 0.5 truecm}

\begin{document}

\title{Using the 6dF galaxy redshift survey to detect gravitationally-lensed
	quasars}

\author{
	Daniel J.\ Mortlock$^{1,2}$ \and
 	Michael J.\ Drinkwater$^{3}$
}

\date{}
\maketitle

{\center
	$^1$ Astrophysics Group, Cavendish Laboratory, Madingley Road,
	Cambridge CB3 0HE,
        United Kingdom \\
	mortlock@ast.cam.ac.uk\\[3mm]
	$^2$ Institute of Astronomy, Madingley Road, Cambridge CB3 0HA,
	United Kingdom \\[3mm]
	$^3$ School of Physics, University of Melbourne,
	Vic 3010, Australia \\
	mdrinkwa@physics.unimelb.edu.au \\
}

\begin{abstract}
It is possible to detect gravitationally-lensed quasars
spectroscopically if the spectra obtained during galaxy surveys are
searched for the presence of quasar emission lines. The up-coming 6
degree Field (6dF) redshift survey on the United Kingdom Schmidt Telescope 
will involve obtaining $\sim 10^5$ spectra of near-infrared selected
galaxies to a magnitude limit of $K = 13$. Applying previously
developed techniques implies that at least one lens should be
discovered in the 6dF survey, but that as many as ten could be found
if quasars typically have $B_{\rm J} - K \simeq 8$. In this model
there could be up to fifty lensed quasars in the sample, but most of
them could only be detected by infrared spectroscopy.
\end{abstract}


\bigskip

\section{Introduction}
\label{section:intro}

Gravitationally-lensed quasars are very valuable.
Individual lenses can be used to constrain 
the properties of the source,
the distribution of mass in the deflector, 
and, through microlensing, the composition of the deflector.
The frequency of lensing can be used to place limits on the 
cosmological model, and the distribution of 
image separations provides a useful probe of the average deflector
mass.
These and other uses of lenses are well known, but cases of multiple
imaging are still rare ($\sim 50$ known to date). 
This is not due to any lack of effort on the part of observers -- 
most lenses are discovered by dedicated re-observation of 
previously identified high-redshift sources, but the likelihood
of lensing is sufficiently low that imaging of up to $\sim 1000$ fields
is required per lensing event. 

It is also possible to search for lenses spectroscopically,
as opposed to morphologically. 
A small fraction of quasars are multiply-imaged by galaxies sufficiently
nearby to make the composite object appear extended, whilst the 
quasar is bright enough for its emission lines to be apparent in 
the composite spectrum. 
Only one quasar lens -- Q 2237+0305 (Huchra et al.\ 1985) -- has 
so far been discovered in
this manner, but the proximity of the deflector (a redshift 0.04 spiral
galaxy) has allowed a number of unique measurements to be made,
as summarised by Mortlock \& Webster (2000).
Spectroscopic lens surveys will primarily be sensitive to these
particularly useful systems, but 
the restriction that the deflector be so nearby also means that the 
frequency of such lenses is considerably lower than that of the 
general lens population (Kochanek 1992; Mortlock \& Webster 2000).
Thus a dedicated spectroscopic lens survey would be disastrously
inefficient, but galaxy redshift surveys (GRSs)
provide large samples of galaxy spectra as a matter of course.
Such lens surveys are potentially very efficient,
with the only data requirements being follow-up imaging 
of the best spectroscopic candidates.

Kochanek (1992) and Mortlock \& Webster (2000) have already 
investigated the statistical likelihood of finding lensed quasars in 
redshift surveys. The 2 degree Field (2dF) GRS (Folkes et al.\ 1999)
should contain at least ten lenses amongst its $\sim 2.5 \times 10^5$
spectra\footnote{Effort is now underway to find lensed quasars in the 
2dF spectra; this project, along with other spectroscopic lens
searches, are described elsewhere in this issue by
Mortlock, Madgwick \& Lahav (2001).}, 
and the larger Sloan Digital Sky Survey (SDSS; York et al.\ 2000),
with close to $10^6$ galaxies,
may yield up to a hundred spectroscopic lenses.

These results are extended to the near-infrared-selected 6 degree Field
(6dF) GRS here.
The survey is described in Section~\ref{section:grs},
and the lens calculation is presented in Section~\ref{section:lens}. 
The results are summarised in Section~\ref{section:conc}.

\section{The 6dF galaxy redshift survey}
\label{section:grs}

The 6dF instrument is a multi-fibre spectrograph for the 
Anglo-Australian Observatory's 
United Kingdom Schmidt Telescope (UKST).
The largest project planned for
the instrument is the 6dF GRS (e.g., Watson et al.\ 1998, 2000). 
This will involve obtaining the spectra of $\sim 1.2 \times 10^5$
galaxy candidates, as identified in the 
2 Micron All-Sky Survey (2MASS; Jarrett et al.\ 2000),
to a limiting magnitude of $K = 13.0$.
About 175 nights of observing time will be required
for the 6dF survey, which should be complete by mid-2003.
The 6dF instrument can obtain 150 spectra simultaneously, but the 
integration time required is rather long, at about an hour, due to
the small aperture of telescope ($\sim 1.5$-m). 
The spectra will cover the range between 3900 \AA\ and 7400 \AA,
with a resolution of 3.5 \AA\ and a signal-to-noise ratio of $\sim 10$
per pixel.

The main scientific justification for the 6dF survey, 
given the existence of the 2dF GRS and the SDSS, is the 
near-infrared selection, as this relates directly to the old
stellar population of the local galaxies.
However there are several other distinctions important to 
the possibility of finding spectroscopic lenses.
With a mean redshift of $\sim 0.05$ the 6dF GRS
is considerably shallower than the 2dF sample,
which has an average redshift of $\sim 0.1$.
The 6dF spectra will have a 
similar signal-to-noise ratio per unit wavelength, but will cover
a slightly smaller range.

\section{Lens statistics}
\label{section:lens}

Mortlock \& Webster (2000) contains a detailed description of how to
calculate the number of spectroscopic lenses expected in a GRS. 
Briefly, one must integrate over the galaxy (i.e., deflector) and 
quasar (i.e., source) populations to find the number of lenses which
satisfy the following three criteria:
\begin{itemize}
\item{The total flux from the composite object (i.e., the sum of the
light from the galaxy and quasar) must be brighter than the survey's
flux limit.}
\item{The quasar must be bright enough to be detectable in the spectrum 
of the lens.\footnote{Note that the 6dF instrument's fibres are
	6 arcsec in diameter, so that there is little distinction 
	between the total magnitude and the fibre magnitude 
	(as defined in Mortlock \& Webster 2001) of an object.
	For spectroscopic lens searches small fibres -- like those of 
	the 2dF instrument -- are desirable as they maximise the 
	relative contribution of the quasar's light to the spectrum.}
Defining the summed quasar images' magnitude as
$m_{\rm q}$ and the galaxy's magnitude as $m_{\rm g}$, this
criterion is taken to be
satisfied if $m_{\rm q} \leq m_{\rm g} + \Delta m_{\rm qg}$
(Kochanek 1992). The value of $\Delta m_{\rm qg}$ depends on the 
quality of the data and the ``distinctiveness'' of the quasars' spectra.}
\item{The galaxy must be sufficiently bright that the lens is 
identified as an extended source; 
it is this condition which ensures that only lenses with nearby
deflectors are selected.}
\end{itemize}

Figure 10 of Mortlock \& Webster (2000) shows how the event rate 
increases with a survey's magnitude limit 
and $\Delta m_{\rm qg}$,
but these results apply to a $B_{\rm J}$-selected GRS,
whereas the 6dF sample is subject to a $K$-band flux limit.
The target list, taken from the 2MASS survey, is also $K$-selected,
but $\Delta m_{\rm qg}$ must be calculated
in one of the optical bands (e.g., $B$, $B_{\rm J}$ $V$, $R$, etc.) 
covered by the 6dF spectra. 
A further complication is the paucity of available information about the
quasar population at infrared wavelengths. 
These difficulties were circumvented by performing the calculation
in the $B_{\rm J}$-band (in order to facilitate comparison with
Mortlock \& Webster 2000), 
parameterising the deflector and source populations
in terms of their $B_{\rm J} - K$ colours.

The $B_{\rm J}$- and $K$-band galaxy luminosity functions 
measured, respectively,
by Folkes et al.\ (1999) and Loveday (2000)
are consistent provided the local galaxy population has
$\langle B_{\rm J} - K \rangle \simeq 4$. 
Hence the 6dF GRS should be approximately equivalent to 
$B_{\rm J}$-selected survey with a magnitude limit of $\sim 17$.

There have been no systematic $K$-band quasar surveys, but several
attempts have been made to infer the luminosity function at these
wavelengths from observations in other bands.  Most quasar samples are
selected using ultraviolet excess (UVX) methods (e.g., Boyle, Shanks
\& Peterson 1988), but dust obscuration at these wavelengths may
result in serious incompleteness.  UVX-selected samples typically have
$\langle B_{\rm J} - K \rangle \simeq 2.5$, but the radio-selected
Parkes Half-Jansky Flat-Spectrum sources (Drinkwater et al.\ 1997) are
considerably redder, with $2 < B_{\rm J} - K < 10$ (Webster et al.\
1995).  These discrepancies should be resolved by upcoming infrared
surveys (e.g., Warren, Hewett \& Foltz 2000), but, for the moment, a
range of $\langle B_{\rm J} - K \rangle$ values must be considered.

The uncertainty in the quasars' colours also affects the determination 
of $\Delta m_{\rm qg}$, their spectral prominence in the survey data. 
The 6dF spectra will be of similar quality to those obtained during the 
2dF survey, for which Mortlock \& Webster (2001) estimated 
$\Delta m_{\rm qg} \simeq 2$. However this figure is relevant only
if UVX-selected quasars are representative of the population as a
whole. Whilst $\Delta m_{\rm qg}$ is calculated in the $B_{\rm J}$-band,
the 6dF spectra extend past $V$ and $R$, almost to $I$, and quasars
with $B_{\rm J} - K \simeq 6$ would be much more prominent in the
red end (close to the $I$-band) of the spectra than UVX-selected objects. 
In terms of the model used here, this leads to an increase in
$\Delta m_{\rm qg}$ by up to a magnitude, although the change depends 
on the exact shape of the galaxy and quasar spectra (specifically,
their $B_{\rm J} - I$ colours).

\begin{table}
\begin{center}
\caption{Number of lenses expected in the 6dF GRS}
\begin{tabular}{ccccc}
\\
\hline
& \multicolumn{2}{c}{optical spectroscopy} 
& \multicolumn{2}{c}{infrared spectroscopy} \\
$\langle B_{\rm J} - K\rangle_{\rm quasar}$ 
& $\Delta m_{\rm qg}$ & $N_{\rm lens}$
& $\Delta m_{\rm qg}$ & $N_{\rm lens}$ \\
\hline 
2 & 1.5 & 0.3 & 0.0 & 0.02 \\
4 & 2.0 & 1 & 2.0 & 1 \\
6 & 2.5 & 3 & 4.0 & 14 \\
8 & 3.0 & 7 & 6.0 & 27 \\
\hline
\label{table:n_lens}
\end{tabular}
\end{center}
\end{table}

The results of integrating over the deflector and source populations
are given in Table~\ref{table:n_lens}. With optical spectroscopy
the 6dF GRS should contain about 5 lenses if quasars 
are as red as inferred by Webster et al.\ (1995), but there
would also be a large number of lensed quasars in the sample that are too red 
to be detected in the optical spectra. 
The only way to find these objects would be to obtain infrared 
spectroscopy of the entire sample; the likely yield from this 
procedure is also given in Table~\ref{table:n_lens}. In this 
case the increase of $N_{\rm lens}$ with $\Delta m_{\rm qg}$ is
not as marked, as such a survey would probe magnitudes at 
which the quasar luminosity function was much flatter
(e.g., Boyle et al.\ 1988)

\section{Conclusions}
\label{section:conc}

Kochanek (1992) and Mortlock \& Webster (2000) showed that several
tens of lensed quasars could be discovered in galaxy surveys
by examining the spectra obtained for the presence of quasar 
emission features. 
The 6dF GRS should contain at least one lens, but could yield as many as
seven if quasars are typically very red (e.g., $\langle B_{\rm J} - K \rangle
\simeq 8$; c.f.\ Webster et al.\ 1995). In this latter case the 
survey will in fact contain several tens of lenses, but most 
of these would only be detectable with infrared spectroscopy.

\section*{Acknowledgements}

DJM is funded by PPARC and MJD is funded by the Australian Research Council.

\section*{References}

\reference Boyle, B.J., Shanks, T., Peterson, B.A., 1988, MNRAS, 235, 935
\reference Drinkwater, M.J., et al., 1997, MNRAS, 284, 85
\reference Folkes, S.R., et al., 1999, MNRAS, 308, 459
\reference Huchra, J.P., Gorenstien, M., Kent, S., Shapiro, I., Smith, G.,
        Horine, E., Perley, R., 1985, AJ, 90, 691
\reference Jarrett, T.H., et al., 2000, AJ, 119, 2498
\reference Kochanek, C.S., 1992, ApJ, 397, 381
\reference Loveday, J., 2000, MNRAS, 312, 557
\reference Mortlock, D.J., Webster, R.L., 2000, MNRAS, 319, 879
\reference Mortlock, D.J., Webster, R.L., 2001, MNRAS, 321, 629
\reference Mortlock, D.J., Madgwick, D.S., Lahav, O., 2001, PASA, in press
\reference Warren, S.J., Hewett, P.C., Foltz, C.B., 2000, MNRAS, 312, 827
\reference Watson, F.G., Parker, Q.A., Miziarski, S., 1998, in
	Optical Astronomical Instrumentation,
	ed.\ D'Odorico, S.,
	SPIE, 834
\reference Watson, F.G., Parker, Q.A., Bogatu, G., Farrell, T.J.,
 	Hingley, B.E., Miziarski, S., 2000, in
	Optical and IR Telescope Instrumentation and Detectors,
	eds.\ Iye, M., Moorwood, A.F.,
	SPIE, 123
\reference Webster, R.L., Francis, P.J., Peterson, B.A., Drinkwater, M.J.,
        Masci, F.J., 1995, Nature, 375, 469
\reference York, D.G., et al., 2000, AJ, 120, 1579 

\end{document}